\documentclass[amsmath,amssymb,aps,twocolumn,notitlepage,showpacs,pre,floatfix,longbibliography]{revtex4-1}

\usepackage{graphicx}
\usepackage{bm}
\usepackage{hyperref}
\usepackage[usenames, dvipsnames]{color}
\usepackage{tabto,lipsum} 
\usepackage{floatrow} 

\hypersetup{
    colorlinks=true,
    linkcolor=blue,
    citecolor=blue,
    filecolor=blue,
    urlcolor=blue,
}
\usepackage[normalem]{ulem}

\begin{document}

\title{Two-temperature Brownian dynamics of a particle in a confining potential}

\author{Vincent Mancois$^{1,2,3}$, Bruno Marcos$^{4}$, Pascal Viot$^{3,1}$, and David Wilkowski$^{1,2,5}$}
\affiliation{$^1$ MajuLab, CNRS-Universit\'e de Nice-NUS-NTU International
Joint Research Unit UMI 3654, Singapore}
\affiliation{$^2$ PAP, School of   Physical and Mathematical Sciences, Nanyang Technological University, 637371 Singapore}
\affiliation{$^3$ Laboratoire de Physique
  Th\'eorique de la Mati\`ere Condens\'ee, Sorbonne Universit\'e, CNRS  UMR 7600,  4, place Jussieu, 75252 Paris Cedex 05, France}
\affiliation{$^4$ Universit\'e C\^ote d'Azur, CNRS, LJAD, 06108 Nice, France}
\affiliation{$^5$ Centre for Quantum Technologies, National University of Singapore, 117543 Singapore}

\begin{abstract}
We consider  the two dimensional  motion of a  particle  into a confining potential,
subjected to 
Brownian forces, associated with two different temperatures on the orthogonal  directions.
Exact solutions are obtained  for an asymmetric harmonic potential in the  overdamped and underdamped regimes, 
whereas perturbative approaches are used for more general potentials.
The resulting non equilibrium stationary state is characterized  with a nonzero  orthoradial mean current, 
corresponding to a global rotation of the particle around the center.
The rotation is due to two symmetry breaking: two different temperatures and a mismatch between the principal axes of
the confining asymmetric potential and the temperature axes.
We confirm our predictions by performing Brownian dynamics simulation.  
Finally, we propose to observe this effect on a laser cooled atomic system.

\end{abstract}
\date{\today}
\maketitle
\section{Introduction}
When a system is in contact with two reservoirs of  different
temperatures or of different chemical potentials, the system does not relax to equilibrium
but is driven toward a non equilibrium stationary state.
Moreover, for small systems, observables are characterized not only by the mean value, but also by fluctuations.
Brownian particles driven by an external force provide paradigmatic models for studying fluctuation theorems and stochastic thermodynamics. When two heat reservoirs
are in contact with a system, a conversion of fluctuations into  directed transport  and also useful work can be observed.
Derrida and
Brunet \cite{derrida2005b} considered a one-dimensional stochastic model in contact with two thermostats describing
the time evolution of a hard rod whose extremities are in contact with two different thermostats.
Visco \cite{Visco2006} obtained the exact large deviation function of the work
fluctuations of the model, showing that the Fluctuation relation has a finite range of validity. (See also \cite{Fogedby2011,Fogedby2014} for some generalizations.)
Van den Broeck {\it et al.} \cite{VandenBroeck2004} proposed an underdamped version of the Derrida and Brunet model and their analysis
revealed that it contains the fundamental building blocks for thermal Brownian motors.
(For a reviews on Brownian motors, see \cite{Broeck2005,Reimann2002}).
More recently Murashita and Esposito \cite{Murashita2016}  have revisited that one-dimensional stochastic models with multiple thermostats in the overdamped limit. They showed 
that it must be carefully considered, as one cannot simply derive the overdamped limit from the underdamped regime. Conversely, matching each reservoir to an independent degree
of freedom of the particle, should ensure  a proper overdamped limit. This later situation, being illustrated for example by two coupled particles in contact with two heat reservoirs,
was experimentally and theoretically studied recently: Ciliberto
{\it et al.} \cite{1742-5468-2013-12-P12014,Ciliberto2013} measured the energy exchanged between two conductors kept at two different
temperatures and linearly coupled. They have analyzed experimental results in terms of two Brownian particles kept at different
temperatures and coupled by an elastic force. B\'erut {\it et al.} \cite{Berut2014,Berut2016} measured the energy flux, the correlation functions and the probability distribution functions of
a system of two particles
in optical traps   with an hydrodynamic coupling at two different
temperatures. Finally, primacy of the coupling strength between particles in minimal thermal motors was demonstrated in \cite{Fogedby2017}. (See also the recent review on experiments in stochastic thermodynamics \cite{Ciliberto2017}).

In this paper, we consider a two-dimensional motion of a particle subjected to two orthogonal Brownian forces of different temperatures denoted
$T_x$ and  $T_y$, respectively.
This particle of mass $m$  is also subjected  to an external conservative force. The overdamped version of this model, in the case of an asymmetric harmonic potential, was previously studied
by Dotsenko et al. \cite{Dotsenko2013}. They derived the non equilibrium probability distribution function (PDF) of positions and
showed the presence of a non-zero current as long as the principal axes of the potential do not coincide with the temperature axes.
We show that the emergence of this current is more general and associated with the altogether two broken symmetries:
two different temperatures in two orthogonal directions and a mismatch between the temperature axes and the principal axes of the potential. Furthermore, we demonstrate that the macroscopic 
rotation is still present for an overdamped Brownian motion and for general confining potentials.

The article is organized as follows. In  Sec. II,  we consider the motion of a particle in an anisotropic harmonic trap for which we obtain exact expressions for PDF of positions and velocities 
as well as for the mean current. This latter being characterized by a mean angular velocity. In Sec. III,  we address the general potential cases, performing a perturbative expansion with 
respect of a small temperature difference $(T_y-T_x)$ and a small asymmetry of the confining potential $U(x,y)$, characterized by a dimensionless parameter $u$,
we show that the mean current velocity is proportional to $u(T_y-T_x)$ at the lowest order. In Sec. IV, we perform numerical simulations of these models, confirming exact solutions 
obtained for a harmonic potential and validating the perturbative approach for the general confining potential. In Sec V., we discuss possible experimental realization on a laser-cooled 
atomic gas. The rotation could be simply observed using standard time of flight (TOF) techniques.

\section{Harmonic potential: exact solutions}

We consider a particle of mass $m$ moving in a  plane. This particle is subjected to a conservative force
deriving from a confining potential $U(x,y)$, a viscous linear force $-\eta {\bf v}$ and two stochastic forces $\sqrt{2T_i\eta}\xi_i(t)$. Here, $\eta$ is the constant friction coefficient,
$T_i$ are the temperatures (expressed in energy unit) along the $i$ axis with $i=x,y$, and $\xi_i(t)$ are uncorrelated Gaussian white noises:
$\langle \xi_i (t)\rangle=0$ and
$\langle \xi_i (t')\xi_j (t)\rangle=\delta(t-t') \delta_{ij}$. Where $\delta(t)$ is the Dirac distribution and $\delta_{ij}$ is the Kronecker symbol. 

\subsection{Overdamped motion}
\subsubsection{Model A}

We now assume that $U(x,y)=k(\frac{x^2+y^2}{2}+uxy)$ (with $|u|<1$ for having a confining potential),
where $k$ is the elasticity constant and $u$ a dimensionless parameter characterizing the potential anisotropy. More precisely, the principal axes of the potential are rotated by $\pi/4$ with respect to $\hat{x}$ and $\hat{y}$, the temperature axes, and $\sqrt{2u/(1-u)}$ is the eccentricity of the iso-potentials.

We  first consider the overdamped motion  in order to introduce the method used to solve the harmonic case \cite{Dotsenko2013}. Then the time evolution is given by the equations
\begin{align}
  \frac{dx}{dt}&=-\frac{k}{\eta}(x+uy)+\sqrt{\frac{2 T_x}{\eta}}\xi_x(t),\nonumber\\
  \frac{dy}{dt} &=-\frac{k}{\eta}(y+ux)+\sqrt{\frac{2 T_y}{\eta}}\xi_y(t).
 \end{align}
The associated Fokker-Planck equation is given by

\begin{align}\label{eq:fokker}
 \frac{\partial P(x,y,t)}{\partial t}&=-\bm{\nabla} \cdot \textbf{J}
 \end{align}
 where $\bm{\nabla}=(\partial_x,\partial_y)$ and

\begin{equation}\label{eq:flux}
 {\bf J}=\begin{cases}
          J_{x}&=-\frac{k}{\eta}(x+uy)P(x,y)-\frac{T_x}{\eta}\frac{\partial P}{\partial x}\\
          J_{y}&=-\frac{k}{\eta}(y+ux)P(x,y) -\frac{T_y}{\eta}\frac{\partial P}{\partial y}.\\
         \end{cases}
\end{equation}
To access the stationary position PDFs , we follow the method detailed in Appendix \ref{Method}. For this purpose,
we introduce the $2 \times 2$ matrices $A$, $B$ and $\Xi$

\begin{equation}
 A=-\frac{k}{\eta}\left(\begin{array}{cc}
          1&u\\u&1
         \end{array}
         \right)
,
 B=\frac{2}{\eta}\left(\begin{array}{cc}
          T_x&0\\0&T_y
         \end{array}
         \right).
\end{equation}

\begin{equation}
 \Xi=\frac{1}{2k(1-u^2)}\left(\begin{array}{cc}
 2T_x+(T_y-T_x)u^2 &-(T_x+T_y)u\\
 -(T_x+T_y)u& 2T_y+(T_x-T_y)u^2
\end{array}
\right).
\end{equation}
The solution to Eq. (\ref{eq:fokker}) is the multivariate Gaussian distribution
\begin{equation}
 P(\textbf{z})=\frac{1}{2\pi \sqrt{Det(\Xi)}}\exp\left(-\frac{1}{2}(\textbf{z}-\langle \textbf{z}\rangle)^T\Xi^{-1}
 (\textbf{z}-\langle \textbf{z}\rangle)\right),
\end{equation}
where $\textbf{z}$ is a two-dimensional vector of components $(x,y)$, $\textbf{z}^T$ its transpose vector and $\langle \textbf{z}\rangle$ its statistical average. Finally, the stationary PDF reads

\begin{equation}\label{eq:fullproba}
 P(x,y)=\frac{k\sqrt{1-u^2}
 e^{-(\gamma_1  x ^2+\gamma_2 y^2+\gamma_3 xy)}}{\pi  \sqrt{4T_xT_y+u^2(T_y-T_x)^2}},
\end{equation}
where
\begin{align}
 \gamma_1&=k\frac{2T_y+u^2(T_x-T_y)}{4T_x T_y+u^2(T_x-T_y)^2},\\
 \gamma_2&=k\frac{2T_x+u^2(T_y-T_x)}{4T_x T_y+u^2(T_x-T_y)^2},\\
 \gamma_3&=k\frac{2u (T_x+T_y)}{4T_x T_y+u^2(T_x-T_y)^2}.
\end{align}

\noindent When $u=0$, one obtains   $P(x,y)\sim e^{-\frac{kx^2}{2T_x}-\frac{ky^2}{2 T_y}}$, which corresponds to two decoupled oscillators at equilibrium.
Integrating Eq. (\ref{eq:fullproba}) over $x$ or $y$, one obtains the marginal distributions $P(x)$ and $P(y)$, respectively, which  have a Gaussian profile.
The variances $\langle x^2\rangle$ and $\langle y^2\rangle$ are given by
\begin{align}\label{eq:Txyhf}
 \langle x^2\rangle&=\frac{T_x+\frac{u^2}{2}(T_y-T_x)}{k(1-u^2)},\\
 \langle y^2\rangle&=\frac{T_y+\frac{u^2}{2}(T_x-T_y)}{k(1-u^2)},
\end{align}
\noindent with a cross-correlation 
\begin{align}
 \langle xy\rangle=-\frac{u(T_x+T_y)}{2k(1-u^2)}.
\end{align}
This last term is non zero only if $u\neq0$.

The non equilibrium stationary state is also characterized by a non-zero
current probability ${\bf J}={\bf v}P$ \cite{Seifert2012}. The  angular velocity is defined as
\begin{equation}
 \omega(t)=\frac{1}{r^2}({\bf r}\times {\bf v}).
\end{equation}
The mean angular velocity is given by the long-time limit
\begin{equation}
 \langle \omega\rangle=\lim_{t\rightarrow\infty}\frac{1}{t}\int_0^t dt' \omega(t').
\end{equation}
Assuming ergodicity of the system, the time average is equivalent to the ensemble average, one has 
\begin{equation}\label{eq:omega}
 \langle \omega\rangle=\int d^2{\bf r} \frac{1}{r^2}( \textbf{r} \times \textbf{v})P({\bf r}).
\end{equation}
Using the polar coordinates, one obtains that
\begin{equation}\label{eq:omegagen}
 \langle \omega\rangle=\int_0^{2\pi}d\theta \int_0^\infty d r  J_\theta(r,\theta),
\end{equation}
with
\begin{align}\label{eq:current}
 J_\theta(r,\theta)&=-\frac{kur \cos(2\theta )P}{\eta}-\frac{1}{2\eta r}(T_x+ T_y)
 \frac{\partial P}{\partial \theta}\nonumber\\
 &-\frac{T_y- T_x}{2\eta }\left[\frac{\cos(2\theta)}{r}
 \frac{\partial P}{\partial \theta}+\sin(2\theta)
 \frac{\partial P}{\partial r}\right].
\end{align}
In polar coordinates, the stationary probability distribution is given by
\begin{equation}\label{eq:Ppolar}
 P(r,\theta)=\frac{k\sqrt{1-u^2}
 e^{-(\gamma_+ +\gamma_- \cos(2\theta)+u\gamma_+ \sin(2\theta))r^2}}{\pi \sqrt{4T_xT_y+u^2(T_y-T_x)^2}},
\end{equation}
with
\begin{align}
 \gamma_+&=k\frac{T_x+T_y}{4T_x T_y+u^2(T_x-T_y)^2},\\
 \gamma_-&=k\frac{(1-u^2)(T_y-T_x)}{4T_x T_y+u^2(T_x-T_y)^2}.
\end{align}
Inserting Eq.(\ref{eq:Ppolar}) in Eq.(\ref{eq:current}), one obtains
\begin{align}\label{jtheta}
 J_\theta(r,\theta)&= \frac{u(T_y-T_x)r}{\eta}\nonumber\\
 &(\gamma_+ +\gamma_- \cos(2\theta)+u\gamma_+ \sin(2\theta))P(r,\theta).
\end{align}
Finally, the mean angular velocity is given by
\begin{equation}\label{eq:omegaoverdamped}
  \langle \omega\rangle=\frac{k}{\eta}u(T_y-T_x)\sqrt{\frac{1-u^2}{4T_xT_y+u^2(T_x-T_y)^2}}.
\end{equation}
The other moments of $  \langle \omega\rangle$ can be also obtained
\begin{align}\label{eq:higheromega}
  \langle \omega^n\rangle&=k\left(\frac{u(T_y-T_x)}{\eta}\right)^n\sqrt{\frac{1-u^2}{4T_xT_y+u^2(T_x-T_y)^2}}\nonumber\\
  &\int_0^{2\pi}  \frac{d\theta}{2\pi} (\gamma_+ +\gamma_- \cos(2\theta)+u\gamma_+ \sin(2\theta)))^{n-1}
\end{align}
which gives a variance
\begin{align}
 \langle \omega^2\rangle-\langle \omega\rangle^2 &=k\left(\frac{u(T_y-T_x)}{\eta}\right)^2 
 \frac{1-u^2}{4T_xT_y+u^2(T_x-T_y)^2}\nonumber\\
&\left[\frac{T_x+T_y}{\sqrt{4T_x T_y+u^2(T_x-T_y)^2}}-1\right].
\end{align}
The variance and all moments also vanish when $u=0$ or $T_y-T_x=0$, which means a total disappearance of a global rotation when the two symmetries are not broken.

\begin{figure}[t!]
 \begin{center}
\includegraphics[scale=0.33]{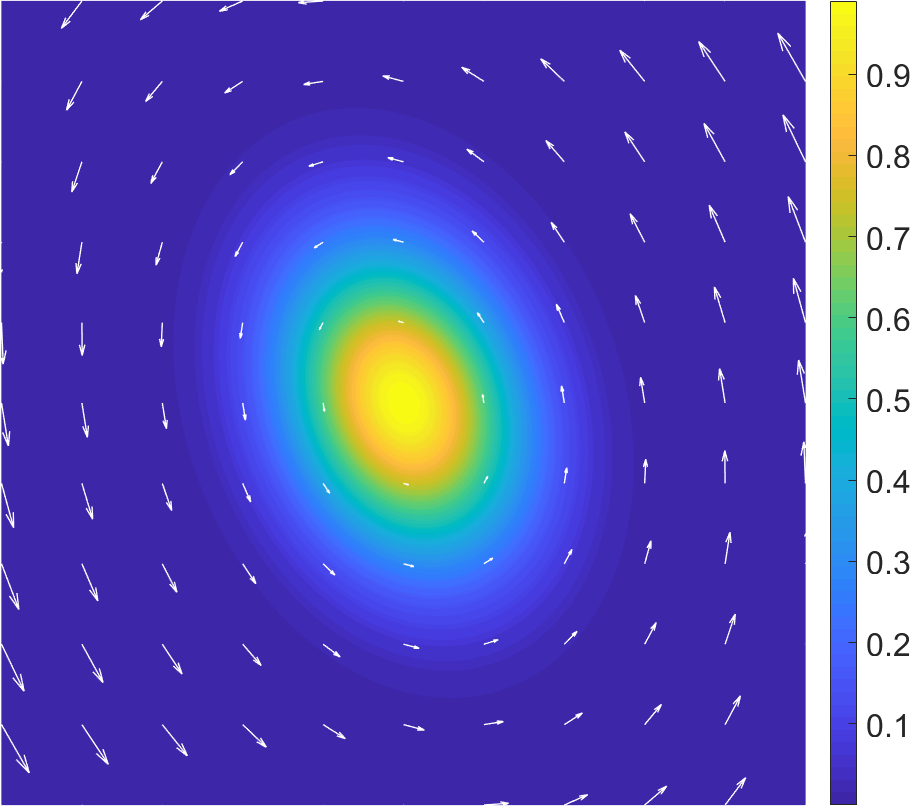}
 \end{center}
 \caption{Density plot of  $P(x,y)$ and  white arrows representing the vector field of the velocity for a harmonic potential with $u=0.2$ and $T_y=2T_x=1$. Color scale darkens toward low density.}
 \label{fig:vfield}
 \end{figure}

Figure \ref{fig:vfield} 
displays a density plot of $P(x,y)$ and the white arrows the vector field of the velocity for a harmonic potential with $u=0.2$ and $T_y= 2T_x=1$. 
The particle density has a maximum in the center whereas the velocity increases linearly with the distance 
the particle to center.

We see that a non-zero current requires  two broken symmetries:  two different temperatures along  the  orthogonal axes which do not match the principal axes of the potential.

\subsubsection{Model B}
In order to show that the non-zero current is not originated from the anisotropy of the potential, but from the mismatch between the principal axes of the potential and the temperature axes, one considers a second model where the particle
evolves in a potential $U(x,y)=k(\frac{x^2+a^2y^2}{2}+uxy)$. Stable potentials require $|a|>|u|$. By using the same method, one obtains an exact solution of  the stationary probability $P(x,y)$, which reads
\begin{equation}\label{eq:fullproba2}
 P(x,y)=\frac{k(1+a^2)\sqrt{a^2-u^2}
 e^{-(\gamma'_1  x ^2+\gamma'_2 y^2+\gamma'_3 xy)}}{\pi  \sqrt{(1+a^2)^2T_xT_y+u^2(T_y-T_x)^2}},
\end{equation}
where
\begin{align}
 \gamma'_1&=k\frac{(1+a^2)(T_y(1+a^2)+u^2(T_x-T_y))}{2((1+a^2)^2T_x T_y+u^2(T_x-T_y)^2)}\\
 \gamma'_2&=k\frac{(1+a^2)(2T_x+u^2(T_y-T_x))}{2((1+a^2)^2T_x T_y+u^2(T_x-T_y)^2)}\\
 \gamma'_3&=k\frac{(1+a^2)u (a^2T_x+T_y)}{(1+a^2)^2T_x T_y+u^2(T_x-T_y)^2}.
\end{align}
Inserting Eq. (\ref{eq:fullproba2}) in Eq. (\ref{eq:omegagen}), one obtains the mean angular velocity:
\begin{equation}\label{eq:omega_anisotrope}
  \langle \omega\rangle=\frac{k}{\eta}u(T_y-T_x)\sqrt{\frac{a^2-u^2}{(1+a^2)^2T_xT_y+u^2(T_x-T_y)^2}}.
\end{equation}

For an anisotropic potential where the confinement is different along the two temperature axes, it is noticeable
that a non-zero mean angular velocity is proportional
to the product $u(T_y-T_x)$, which means that only the temperature difference and  the part of the anisotropy of the potential  outside of the temperature axes
are relevant. As Eq.~\eqref{eq:omega_anisotrope} displays it prominently, the effect of the anisotropy along the $y$-axis modifies slightly the amplitude of angular velocity. Nevertheless, this trap anisotropy only dresses the effect arising from the double symmetry breaking.


\subsection{Underdamped motion}
Backtracking to the situation of an anisotropic potential $U(x,y)=k(\frac{x^2+y^2}{2}+uxy)$ and considering an underdamped motion (see also \cite{Crisanti2012}), the dynamical equations of a particle are given by
\begin{align}
  \frac{dv_x}{dt}&=-\frac{1}{m}\frac{\partial U(x,y)}{\partial x}-\frac{\eta}{m}  v_x+\sqrt{\frac{2\eta T_x}{m^2}}\xi_x(t),\nonumber\\
  \frac{dv_y}{dt}&=-\frac{1}{m}\frac{\partial U(x,y)}{\partial y}-\frac{\eta}{m}  v_y+\sqrt{\frac{2\eta T_y}{m^2}}\xi_y(t),\nonumber\\
 \frac{dx}{dt}&=v_x,\nonumber\\
 \frac{dy}{dt}&=v_y.
 \end{align}
The Kramers-Fokker-Planck equation corresponding to the underdamped motion is given by

\begin{align}
 \frac{\partial P(x,y,p_x,p_y,t)}{\partial t}&=-\bm{\nabla} \cdot \textbf{J},
 \end{align}
 where $\bm{\nabla}=(\partial_x,\partial_y,\partial_{v_x}\partial_{v_y})$ and
\begin{equation}
 \textbf{J}=\begin{cases}
          J_x&=v_xP\\
          J_y&=v_yP\\
          J_{v_x}&=\left(-\frac{1}{m}\frac{\partial U(x,y)}{\partial x}-\eta\frac{v_x}{m}\right)P
          -\frac{\partial}{\partial v_x}(\frac{\eta T_x}{m^2} P)\\
          J_{v_y}&=\left(-\frac{1}{m}\frac{\partial U(x,y)}{\partial y}-\eta\frac{v_y}{m}\right)P
          -\frac{\partial}{\partial v_y}((\frac{\eta T_y}{m^2}  P).
         \end{cases}
\end{equation}
The stationary PDF  depends now on the variables $x$, $y$, $v_x$ and $v_y$ which are defined as components of
a $4$-component vector $\textbf{z}$. The associated matrices $A$ and $B$ are given by

\begin{equation}
 A=\frac{1}{m}\left(\begin{array}{cccc}
          0&0 &m  &0\\
          0&0 &0 &m\\
	  -k&-ku&-\eta&0\\
	  -ku&-k&0&-\eta
         \end{array}
         \right),
 B=\frac{\eta}{m^2}\left(\begin{array}{cccc}
          0&0&0&0\\
          0&0&0&0\\
          0&0&2 T_x&0\\
          0&0&0&2 T_y
         \end{array}
         \right).
\end{equation}

\begin{figure}[t!]
 \begin{center}
\includegraphics[scale=0.40]{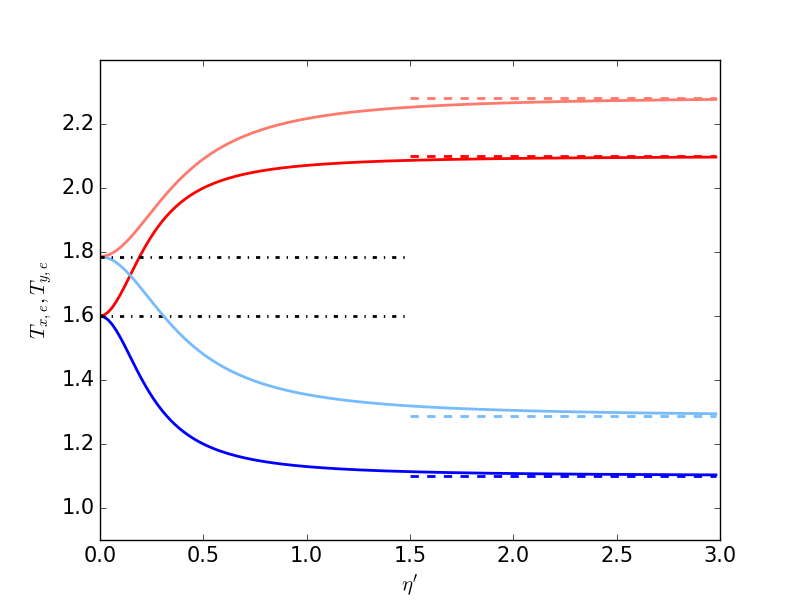}
 \end{center}
 \caption{Effective temperatures $T_{xe}$ and $T_{ye}$ of the position PDFs versus the viscosity $\eta$ for
 $T_x=1$, $T_y=2$ and $u=0.25$ (lower curves) and $u=0.4$ (upper curves). 
 The plain lines correspond to the temperatures along the directions $Ox$ (blue) and $Oy$ (red). The dashed lines give the high friction limit, where the two effective temperatures are larger and smaller than $T_x$ and $T_y$. The dashed-dotted lines indicate the zero friction limit, respectively.}
 \label{fig:tempxy}
 \end{figure}
Introducing the dimensionless viscosity  $\eta'=\eta/\sqrt{km}$ and solving Eq.~(\ref{eq:matrix}), one obtains
\begin{widetext}
\begin{equation}
 \Xi=
 \left( \begin {array}{cccc}
 {\frac { 2\eta'^2 T_x+u^2(T_x+T_y)+\eta'^2(T_y -T_x)}{ 2k\left( \eta'^2 +u^2  \right)  \left(1- u^2  \right) }}&
{-\frac {u \left( T_x +T_y  \right) }{2k(1-u^2)}}&0&
-{\frac {\eta'  \left( T_x -T_y  \right) u}{2\sqrt{km}({\eta'}
^{2}+u^2 )}}\\
\frac {-u \left( T_x + T_y \right) }{2k(1-u^2)}&
{\frac { 2\eta'^2 T_y+u^2(T_x+T_y)+\eta'^2(T_x -T_y)}{ 2k\left( \eta'^2 +u^2  \right)  \left(1- u^2 \right) }}&
{\frac {\eta'  \left( T_x -T_y  \right) u
}{2\sqrt{km}(\eta'^2 +u^2 )}}&0\\ 0&{\frac {\eta'
 \left( T_x -T_y  \right) u}{2\sqrt{km}(\eta'^2 +u^2 )}}&{\frac
{2 T_x \eta'^2 +(T_x  +T_y ) u^2 }{2m(\eta'^2 +u^2) }}&0\\ -{\frac {\eta'  \left( T_x -T_y  \right) u}{2\sqrt{km}({\eta'}
^{2}+u^2 )}}&0&0&{\frac {2 T_y  \eta'^2 +(T_x+T_y)  u^2 }{2m(\eta'^2 +u^2 )}
}\end {array} \right).
\end{equation}
\end{widetext}
By calculating the inverse of $\Xi$, one obtains the probability distribution $P(x,y,v_x,v_y,t)$.
The  marginal probability distributions, namely the position PDF,
$P_x, P_y$ and velocity  $P_{v_x}, P_{v_y}$ can be calculated.
One then obtains for the position PDF  $P_x, P_y$
\begin{equation}\label{eq:distrip}
P_{z}(z)=\sqrt{\frac{1}{2\pi T_{ze}}}\exp(-\frac{z^2}{2T_{ze}})
\end{equation}
where $z=x,y$.

$T_{xe}$  is  the effective temperature of $P_x$ given by
\begin{equation}\label{eq:Txe}
 T_{xe}=\frac {\eta'^2T_x+\frac{u^2}{2}\left(\eta'^2(T_y-T_x)+(T_x+T_y)\right)}{(1-u^2)\left(\eta'^2+u^2\right)}.
\end{equation}
Similarly,
$T_{ye}$ is given by
\begin{equation}\label{eq:Tye}
 T_{ye}=\frac {\eta'^2T_y+\frac{u^2}{2}\left(\eta'^2(T_x-T_y)+(T_x+T_y)\right)}{(1-u^2)\left(\eta'^2+u^2\right)}.
\end{equation}
When $\eta'\gg 1$, Eqs. (\ref{eq:Txe}) and (\ref{eq:Tye}) tend to the overdamped limit given by Eqs.~\eqref{eq:Txyhf}.
Conversely, when $\eta' \rightarrow 0$, one obtains $T_{xe}=T_{ye}=\frac{T_x+T_y}{2(1-u^2)}$.

Figure \ref{fig:tempxy} shows  how $T_{xe}$ and $T_{ye}$ depends on the dimensionless viscosity $\eta'$
for $T_x=1$ and $T_y=2$ (full lines). Note that effective temperatures
$T_{xe}$ and $T_{ye}$,
which start from the same value $(T_x+T_y)/2(1-u^2)$, respectively decreases and increases rapidly towards their asymptotic values, $T_{xe}(\infty)$ and $T_{ye}(\infty)$.
In other words, at low viscosity, the width of the position PDF are given by the mean
temperature of both directions (up to a $(1-u^2)^{-1}$ factor), whereas when the dimensionless viscosity increases, the effective temperatures goes
rapidly towards the asymptotic values (which is independent of the viscosity). 
Note the effective temperatures along each direction is different of $T_x$ and $T_y$, respectively.
Another feature, shown in Fig.~\ref{fig:tempxy}, is the interplay between $u$ and $\eta'$. We indeed notice the inflection point, located at $\eta'\approx u/\sqrt{3}$, marking the crossover between overdamped and underdamped regimes.

\begin{figure}[t!]
\begin{center}
\includegraphics[scale=0.40]{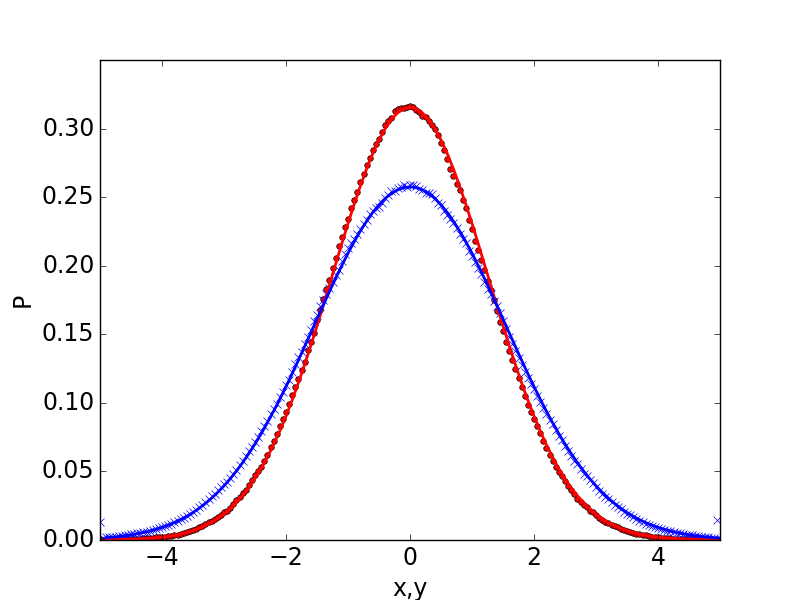}
 \end{center}
 \caption{Stationary position  PDF $P(x)$ (red) and $P(y)$ (blue) for an asymmetric harmonic potential ($u=0.5$), a dimensionless viscosity  $\eta'=1$,
 with   $T_x=1$ and $T_y=2$.}\label{fig:distrip}. 
 \end{figure}
Figure \ref{fig:distrip} displays  the position PDFs, $P_x$ and
$P_y$ for two temperatures $T_x$ and $T_y$ and
$\eta'=1$ (full curves). Dashes curves correspond to the underdamped Langevin simulation (see section \ref{simulations} for more details) and show an accurate agreement
with  the exact expressions, Eq.~\eqref{eq:distrip}.

We now consider the velocity PDF for which one  obtains an exact expression
for the harmonic model. Integrating the position-velocity PDF over the position, it is easy to
show that the velocity PDF are Gaussian with an effective temperature
$T_{v_xe}$ and $T_{v_ye}$ given by

\begin{align}
 T_{v_xe}&=\frac{2T_x\eta'^2+u^2(T_x+T_y)}{2(u^2+\eta'^2)},\\
 T_{v_ye}&=\frac{2T_y\eta'^2+u^2(T_x+T_y)}{2(u^2+\eta'^2)}.
\end{align}
As expected, when $\eta'\rightarrow\infty$, one recovers that $T_{v_xe}=T_x$ and $T_{v_ye}=T_y$ irrespective of  $u$, which means
that the velocity distribution is independent of the potential in the high friction limit. Conversely, when $\eta'\rightarrow 0$,
the two effective temperatures, $T_{v_xe}$ and $T_{v_y,e}$  go to the same limit $(T_x+T_y)/2$.

\begin{figure}[!t]
\begin{center}
\includegraphics[scale=0.40]{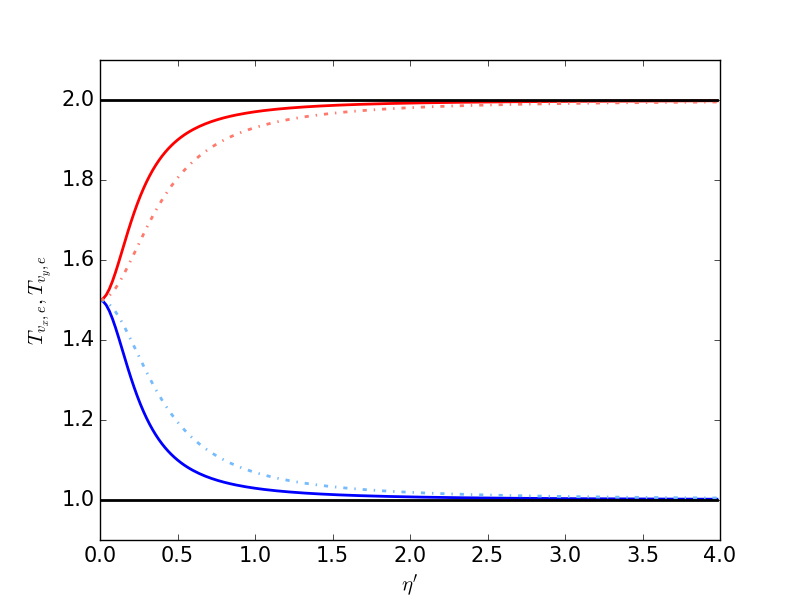}
 \end{center}
 \caption{Effective temperatures $T_{v_x,e}$ and $T_{v_y,e}$ of the velocity PDFs for an asymmetric harmonic potential  with a viscosity  $\eta=1$, $T_x=1$,
 $T_y=2$ and for two values of the asymmetry parameter $u=0.25$ (full curves) and $u=0.4$ (dot-dashed curves). The magenta lines gives the $\eta' \rightarrow + \infty$ limit.}
 \label{fig:tempv}
 \end{figure}

Figure \ref{fig:tempv} shows $T_{v_x,e}$ and $T_{v_y,e}$ as a function of viscosity for two values of the asymmetry parameter $u=0.25,0.4$.
As previously observed for the effective temperatures $T_{x,e}$ and $T_{y,e}$ of the position PDFs, the asymptotic values of
the high friction limit are rapidly reached ($\eta'>2$). However, whereas $T_{x,e}$ and $T_{y,e}$
goes to asymptotic values which depend on the asymmetry parameter $u$ and on the two temperatures
$T_x$ and $T_y$, $T_{v_x,e}$ and $T_{v_y,e}$ goes to $T_x$ and $T_y$, respectively.

Whereas the stationary positions and velocities PDFs are symmetric and has a Gaussian shapes,
the particle motion exhibits a mean rotation velocity. 

\begin{equation}
\label{average-omega}
 \langle \omega\rangle=\int d^2{\bf r} \int d^2{\bf v}\frac{1}{r^2}( {\bf r} \times {\bf v})P({\bf r},{\bf v}).
\end{equation}

By using polar coordinates, the mean angular velocity is expressed as
\begin{equation}
 \langle \omega\rangle=\int dv_r\int dv_\theta \int d\theta \int d r\,  v_\theta  P(r,\theta,v_r,v_\theta).
\end{equation}
After some calculation, one obtains the expression

\begin{align}\label{eq:exactome}
  \langle \omega\rangle&=\frac {k u(T_y-T_x)  \sqrt{ 1-u^2 }}
{\eta\sqrt { (4 T_x\,T_y+ u^2(T_x-T_y)^2)+ \left(\frac{u^4}{\eta'^4}+2\frac{u^2}{\eta'^2}\right)(T_y+T_x)^2} }.
\end{align}

In the overdamped limit ($\eta'\rightarrow\infty$) one recovers Eq.~(\ref{eq:omegaoverdamped}).
In the opposite limit $\eta'\rightarrow 0$, the mean angular velocity decreases to zero as the inverse of the particle mass.

\begin{figure}[t!]
\begin{center}
\includegraphics[scale=0.40]{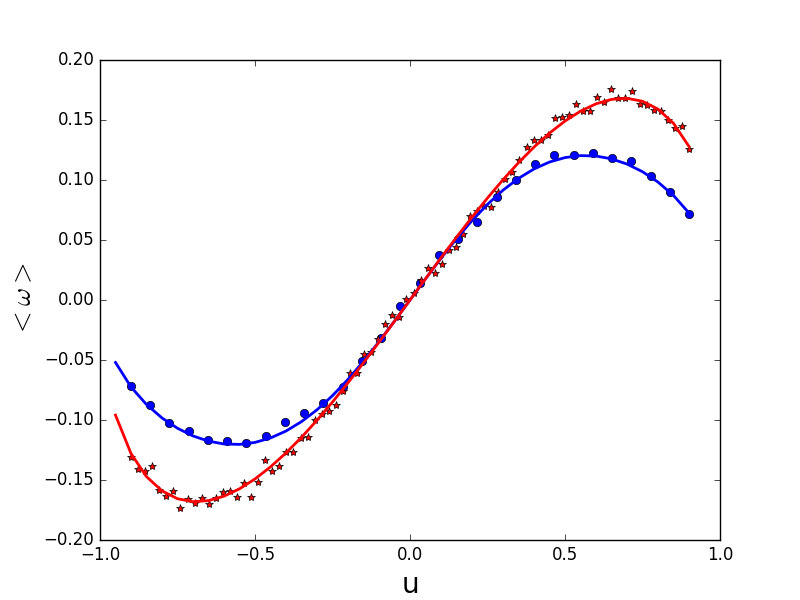}
 \end{center}
 \caption{Numerical simulations of the mean angular velocity  $\langle \omega \rangle$ (in rad.s$^{-1}$) as function of the  asymmetry parameter $u$ for two different  viscosities $\eta'=1,5$, when  $T_x=1$, $T_y=2$ (circles and stars). The full curves  corresponds to the exact expression in the high friction limit.}
 \label{fig:omega}
 \end{figure}

Figure \ref{fig:omega} shows the evolution of $\langle \omega \rangle$ as a function of  the asymmetry parameter $u$ for two values $\eta'=1,5$.
The full curves correspond to Eq. \ref{eq:exactome}. As previously observed with other quantities, for
$ \eta'>3$, the mean angular velocity matches the exact expression corresponding to the high-friction limit.
Note that for a  given value $u$,  the friction coefficient has a weaker impact  than  the asymmetry parameter $u$.

We shown  that the two-temperatures model has a stationary solution with a non zero current
when both the potential is asymmetric and the temperatures are different. One notes that the current is maximum in  the overdamped situation.

\section{Weak asymmetric potential}
\label{weak-asymmetric}
We now consider an overdamped motion of a particle $m$ in a weak asymmetric potential
$U(r,\theta)=U_0(r)+uU(r,\theta)$, where $u\ll 1$ is a small dimensionless parameter and $U_1(r)\leq U_0(r)$ when $r$ is large. Moreover, we consider that the two temperatures are also close, in order to perform the perturbative expansion of the Fokker-Planck equation.
One defines the mean temperature $T=(T_x+T_y)/2$.
The two temperatures along the two axis are
$T_x=T(1-\alpha/2)$ and $T_y=T(1+\alpha/2)$, and $\alpha\ll 1$ is a small dimensionless parameter. 
Using Eq.~\eqref{eq:flux} and expressing the current in polar coordinates, the stationary solution $P(r,\theta)$ satisfies
\begin{equation}\label{eq:divergence}
\frac{1}{r} \frac{\partial (r J_r)}{\partial r}+\frac{1}{r}\frac{\partial J_\theta}{\partial \theta}=0,
\end{equation}
where the radial  and orthoradial currents , $J_r$ and $J_\theta$, are the sum of the two contributions
\begin{equation}
 J_{r,\theta}=J^{1}_{r,\theta}+J^2_{r,\theta},
\end{equation}
where
\begin{align}
 J^1_r(r,\theta)&=-\frac{P}{\eta}\frac{\partial U}{\partial r}-\frac{T}{\eta}\frac{\partial P}{\partial r},\\
 J^1_\theta(r,\theta)&=-\frac{P}{\eta r}\frac{\partial U}{\partial \theta}-\frac{T}{\eta r}\frac{\partial P}{\partial \theta}\label{eq:J1theta}
\end{align}
and
\begin{align}
 J^2_r(r,\theta)&=
  -\frac{\alpha T}{2\eta}\left(-\cos(2\theta)
 \frac{\partial P}{ \partial r}
  +\frac{\sin(2\theta)}{r} \frac{\partial P}{\partial \theta}\right),\\
   J^2_\theta(r,\theta)&=-\frac{\alpha T}{2\eta}\left(\sin(2\theta) \frac{\partial P}{\partial r}+\frac{\cos(2\theta)}{r}
 \frac{\partial P}{\partial \theta} \right).\label{eq:J2theta}
\end{align}
$J^1$ and $J^2$ are the current associated with
the mean temperature and with the temperature difference along the two axis, respectively.
When $\alpha=0$, the stationary solution of the Fokker-Planck equation is the
equilibrium distribution
$P^u_0(r,\theta)\propto e^{-(U_0(r)+u U_1(r,\theta))/T}$, where the associated current vanishes.

In order to perform a perturbative expansion, we propose the following ansatz for the stationary distribution
\begin{equation}\label{eq:ansatz}
 P(r,\theta)= P^u_0(r,\theta)P_1(r,\theta),
\end{equation}
which gives
\begin{align}\label{eq:fluxne}
 J^1_r(r,\theta)&=-\frac{T P_0^u(r,\theta)}{\eta}\frac{\partial P_1 (r,\theta)}{\partial r},\nonumber\\
 J^1_\theta(r,\theta)&=-\frac{T P_0^u(r,\theta)}{\eta r}\frac{\partial P_1(r,\theta)}{\partial \theta}.
\end{align}
Inserting Eq.(\ref{eq:fluxne}) in Eq.(\ref{eq:divergence}), one finally obtains
\begin{equation}
 P^u_0(r,\theta) \Delta P_1(r,\theta) + \vec{\nabla} P^u_0(r,\theta).\vec{\nabla} P_1 (r,\theta) = -\frac{\eta}{T}\vec{\nabla}\cdot\vec{J}^{2}.
\end{equation}
Assuming that $P_1(r)\propto\exp(-\alpha f(r,\theta,u,\alpha))$, and performing a first-order expansion in $\alpha$ (and a zero-order expansion in $u$), one
has
\begin{align}
 \Delta P_1(r,\theta)&=-\alpha P_1(r,\theta)  \Delta  f(r,\theta)+O(\alpha^2),\\
 \vec{\nabla} P^u_0(r,\theta).\vec{\nabla} P_1(r,\theta)&= \frac{\alpha}{T} P^0_0(r)  P_1(r,\theta)
\frac{\partial U_0(r)}{\partial r} \frac{\partial f(r,\theta)}{\partial r}\nonumber\\
&+O(\alpha^2),
\end{align}
and
\begin{align}
 \vec{\nabla}\cdot\vec{J}^{2}&=\frac{\alpha\cos(2\theta)}{2\eta} P^0_0(r)  P_1(r,\theta)\left[\frac{1}{r}\frac{\partial}{\partial r}
 \left(r\frac{\partial U_0(r)}{\partial r}\right)\right.\nonumber\\
 &\left.-\frac{1}{T}\left(\frac{\partial U_0(r)}{\partial r}\right)^2-\frac{2}{r}\frac{\partial U_0(r)}{\partial r}\right]+O(\alpha^2).
\end{align}
Therefore, the function $f(r,\theta)$ satisfies the partial differential equation

\begin{align}
\Delta f(r,\theta)& -\frac{1}{T} \frac{\partial U_0(r)}{\partial r}.\frac{\partial f(r,\theta)}{\partial r}
  =
 \frac{\cos(2\theta)}{2 T} \left[\frac{1}{r}\frac{\partial}{\partial r}
 \left(r\frac{\partial U_0(r)}{\partial r}\right)\right.\nonumber\\
 &\left.-\frac{1}{T}\left(\frac{\partial U_0(r)}{\partial r}\right)^2-\frac{2}{r}\frac{\partial U_0(r)}{\partial r}\right].
\end{align}
Inserting that $f(r,\theta)=\cos(2\theta)g(r)$, on obtains a differential equation for $g(r)$.

\begin{align}\label{eq:gder}
 &\frac{d^2 g(r)}{dr^2}+\frac{1}{r}\frac{dg(r)}{dr}-4\frac{g(r)}{r^2}-\frac{1}{T} \frac{\partial U_0(r)}{\partial r}.\frac{d g(r)}{dr}=\nonumber\\
 &_quad\frac{1}{2 T} \left[\frac{1}{r}\frac{\partial}{\partial r}
 \left(r\frac{\partial U_0(r)}{\partial r}\right)\right.
 \left.-\frac{1}{T}\left(\frac{\partial U_0(r)}{\partial r}\right)^2-\frac{2}{r}\frac{\partial U_0(r)}{\partial r}\right].
\end{align}

The analytic solution of the differential cannot be obtained in general. However, assuming that $U_0(r)\sim r^\alpha$ (with $\alpha\geq 2$) when $r\rightarrow\infty$, one obtains that the asymptotic behavior of $g(r)$ is $g(r)\sim U_0(r)/(2T)$.  A solution exists for the harmonic potential
where $g(r)=kr^2/(4T$). The probability distribution is then given by 
\begin{equation}\label{eq:Pharmonic1}
 P(r,\theta)\propto \exp(-\frac{kr^2}{2T}(1+u\sin(2\theta)+\frac{\alpha}{2} \cos(2\theta)),
\end{equation}
which corresponds to the lowest order expansion in $u$ and $\alpha$ of Eq.~(\ref{eq:Ppolar}) and a mean angular velocity given $\langle \omega \rangle\propto \alpha u kT/\eta$.

Similarly, we now show that the ansatz gives the leading behavior of the mean angular velocity. Using that 
$P(r,\theta)\propto \exp(-(U_0(r)+u U_1(r,\theta))/T-\alpha \cos(2\theta) g(r)))$  Eq.~(\ref{eq:J1theta}) becomes
\begin{equation}
 J^1_\theta(r,\theta)=-\frac{2T\alpha}{\eta r} g(r)\sin(2\theta)P(r,\theta).
\end{equation}

The Fourier series of the anisotropic  part of the potential is  $U_1(r,\theta)=\sum_{n\geq 2}[a_n(r)\cos(n\theta)+b_n(r)\sin(n\theta)]$. 
Because the principal axes  of the potential mismatch the temperature axes, this implies that $b_2(r)$ is nonzero (or at least a  single $b_n(r)$, $n\ge2$, is nonzero). Performing an expansion in $u$ and $\alpha$ of 
Eq.~(\ref{eq:Pharmonic1}), one obtains
that the integral of $J^1(r,\theta)$ over $\theta$ is proportional to $\alpha u T$. 

Similarly, inserting the ansatz of $P(r\theta)$ in Eq.~(\ref{eq:J2theta}), the leading term of the current is given 
by 
\begin{equation}
 J^2_\theta(r,\theta)=+\frac{\alpha }{2\eta}\left(\sin(2\theta)\frac{\partial U_0(r,\theta)}{\partial r}\right)P(r,\theta).
\end{equation}
The integration of $J^2_\theta(r,\theta)$ provides a second contribution of $\langle \omega\rangle$ which  is also 
proportional to $\alpha u T$. Note that when $g(r)=\frac{r}{4T}\frac{\partial U_0(r)}{\partial r}$ the two contributions vanish. For this case which corresponds to the harmonic potential, the orthogonal current must calculated to the next order, which is proportional to $u\alpha$ and one recovers the result obtained in section II, which leads a mean angular velocity proportional to $u\alpha T$. Finally, in all cases, we have shown that the mean angular velocity is proportional to $\alpha u T$, (for $u,\alpha <<1$) which   confirms the fact that the existence of the mean current is
associated with the double symmetry breaking.



\section{Simulation}\label{simulations}

 \begin{figure}[t!]
\begin{center}
\includegraphics[scale=0.25]{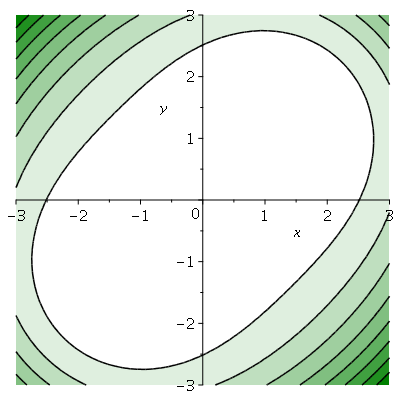}
\includegraphics[scale=0.25]{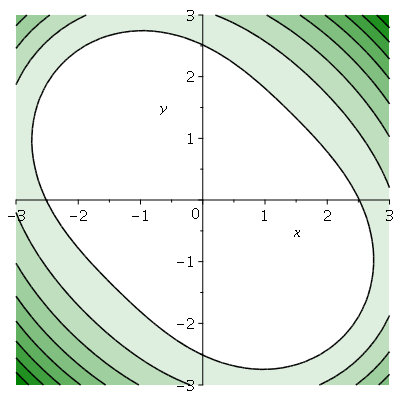}
 \end{center}
 \caption{Contourplot of the potential $U(x,y)=(x^2+y^2)^2/4+uxy$ for $u=3$ (left) and $u=-3$ (right).}
 \label{fig:u4}
 \end{figure}
We performed  stochastic simulation of a particle in the underdamped situation in order to test
the results obtained in the high friction limit. To solve the stochastic differential equation in the underdamped situation, we implement a Verlet-like algorithm
which has the property of using one random number per time step \cite{doi:10.1080/00268976.2012.760055}. Each run is performed  with a total reduced elapsed time $3000$.
Several quantities are monitored in the stationary state: the position probability distributions, the velocity probability distributions. In order
to obtain reliable statistics, one considers the probability distribution along each axis instead of two-dimensional probability distribution.

 \begin{figure}[t!]
\centering
\includegraphics[scale=0.28]{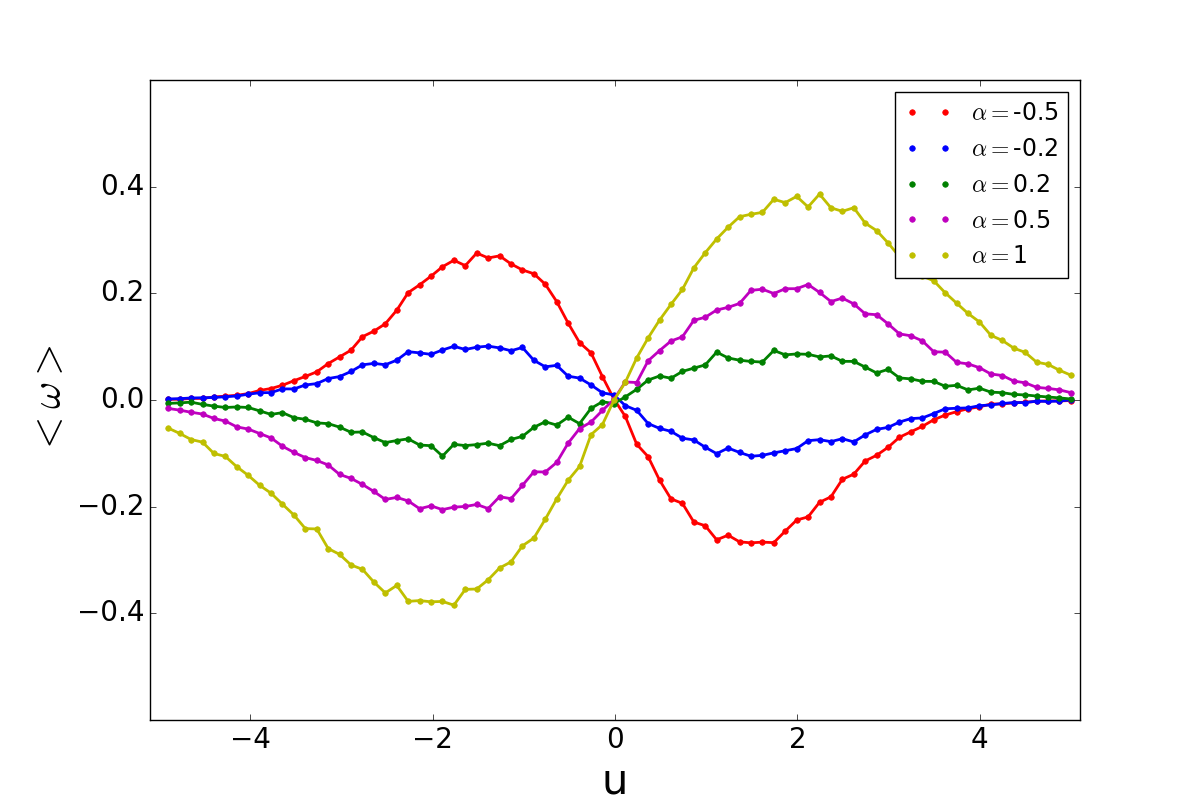}
 \caption{Mean angular velocity versus the asymmetry parameter $u$ for the potential $U(x,y)=(x^2+y^2)^2/4+uxy$. for $T_x=1$ and $T_y=2$.}
 \label{fig:omegaquartic}
 \end{figure}

We first consider the harmonic potential and all observables (position and velocity PDF) mean angular velocity
match the exact results for any value of the viscosity. In particular, we recover the overdamped limit
very rapidly when $\eta'>3$.

\begin{figure}[ht!]
\centering
\includegraphics[scale=0.28]{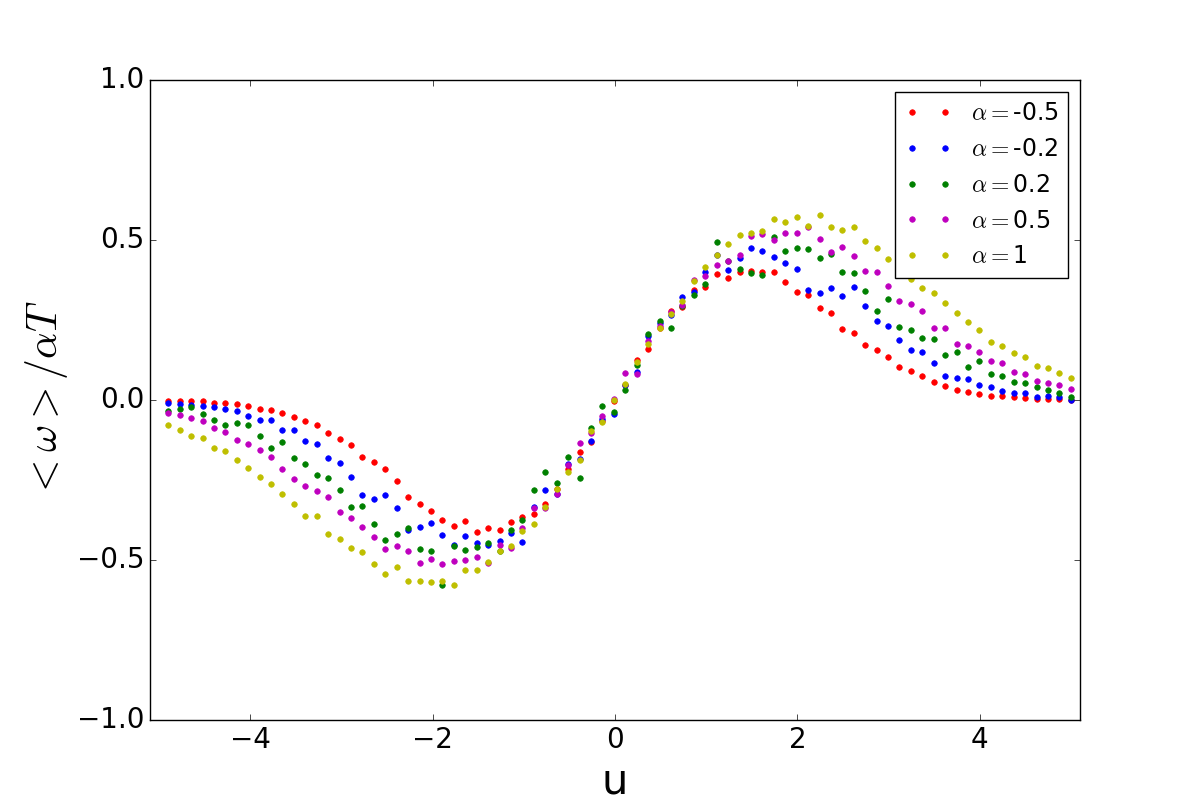}
 \caption{Rescaled mean angular velocity $\omega/( \alpha T)$ versus the asymmetry parameter $u$ for the potential
 $U(x,y)=(x^2+y^2)^2/4+ u xy$ and
 $\alpha=1,0.5,0.2,-0.2,-0.5$.}
 \label{fig:omegaquarticmaster}
 \end{figure}

As seen above, no exact expression is obtained even for a spherical potential except the harmonic potential. We first simulate the model  for $U(x,y)=(x^2+y^2)^2/4+uxy$.
Fig. \ref{fig:u4} displays the asymmetry of the potential for $u=5$ and $u=-5$.
The mean angular velocity is plotted as a function of the asymmetry parameter $u$ (see Fig.\ref{fig:omegaquartic}) for different values of $T_y=2,1.5,1.2,0.8,0.5$  and $T_x=1$. These values corresponds to $\alpha=1,0.5,0.2,-0.2,-0.5$ and
$T=1.5,1.1,0.9,0.75$, respectively. The perturbative analysis of section \ref{weak-asymmetric} predicts that the mean angular velocity is proportional to $\alpha u T$ when $u,\alpha\ll 1$. Fig. \ref{fig:omegaquarticmaster} displays the reduces mean angular velocity $\omega/\alpha T$ as a function
of $u$, and we observe that all the curves collapse for small values of $u$. 
For $u=0$, no mean current exists.  For $u>1.5$, a non linear dependence on $u$ appears and  two extreme values of the mean angular velocity exists for $u\simeq \pm 2$.
In addition, the intensity of the mean angular velocity is  increased compared to the harmonic case.

\section{Observation on cold atoms}

\begin{figure}[!t]
\caption{Time of flight simulations (TOF) emphasizing the difference of behavior when reversing the rotation. The spatial density is plotted after different times of TOF for both clockwise rotation (upper) and counter-clockwise rotation (lower).}\label{fig:tof}
{\includegraphics[scale=0.19]{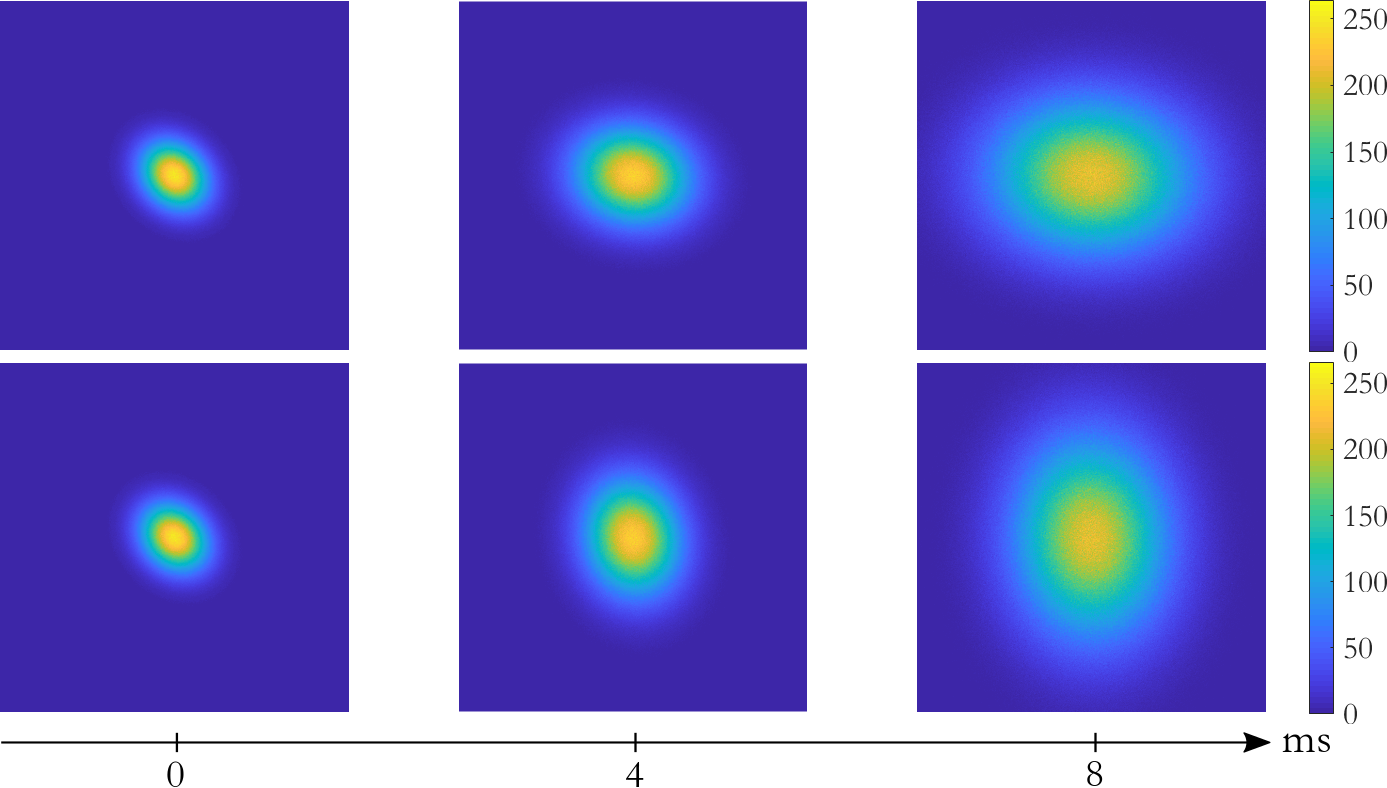}}
\end{figure}

Cold atoms experiments can deal with intensity or laser detuning imbalance during the cooling phase, leading to different temperatures along the different cooling axis. By adding a two-dimensional optical dipole trap, one can as well tailor the asymmetric parameter $u$ at will. Breaking the two symmetries in this case, we wonder to what extent this thermal rotation can be observed. Rotation of atomic clouds have been previously reported in the context of a beam-misaligned vortex trap \cite{Sesko1991,Walker1992} and more recently using synthetic Lorentz forces \cite{Dubcek2014,Santic2015,Santic2017}. As opposed to those previous studies where the rotation is due to a net mean radiation pressure forces, our proposal is based on a stochastic force with zero mean value.

We compute $\langle \omega \rangle$ for optical molasses, in the presence of an optical dipole trap using a semi-classical approach of laser cooling (see for example \cite{Stenholm1986}). For simple experimental implementation, we consider a two-dimensional laser cooling where all laser beams have the same frequency detuning $\delta$ with respect to an atomic transition but with different laser intensities $I_x$ and $I_y$, respectively along the $x$-axis and $y$-axis. We can then define the saturation parameter $s_i=(I_i/I_s)/(1+4\delta^2/\Gamma^2)$, where $I_s$ is the saturation intensity of the atomic transition, $\Gamma$ the atomic linewidth and $i=x,y$. In what follows, we consider the low saturation limit, namely $s_i \ll 1$, so we can sum up the individual contributions of each laser beams to the total radiation pressure force (mean and fluctuating parts). The expansion for the viscous force (mean component) along the $i$-axis reads:

\begin{center}
\begin{equation}\label{radpressure}
\textbf{F}_i=-\eta_i \textbf{v} \qquad \eta_i =-4 \hbar k^2 s_i \frac{2 \delta/\Gamma}{1+(2\delta/\Gamma)^2},
\end{equation}
\end{center}

where $k$ the wavevector of the laser beams and $\hbar$ the Planck constant.

The diffusion constant (fluctuating component), along the $x$-axis reads

\begin{equation}\label{diffcoef}
D_x=\frac{1}{4} \hbar^2 k^2 \Gamma (s_x+s_y) + \frac{1}{2} \hbar^2 k^2 \Gamma s_x.
\end{equation}
The first term on the right side, comes from the photon spontaneous emission events (isotopic radiation pattern), whereas the second term is due to the laser photon absorption events. A similar expression is found along the $y$-axis swapping subscripts $x$ and $x$ in Eq.~\eqref{diffcoef}. Additionally, we assume $\eta=(\eta_x+\eta_y)/2$ to simplify the calculation. More precisely, we could set $\eta_x=\eta_y$ and $D_x\neq D_y$ choosing different frequency detuning and different intensity.

The temperatures along one direction $i=x,y$ are given by the Einstein-Smoluchowski relation:
\begin{center}
\begin{equation}
T_i=\frac{D_i}{\eta}.
\end{equation}
\end{center}
The presence of an asymmetric optical dipole trap $U(x,y)=\frac{m \omega_T^2}{2}(x^2+y^2+2uxy)$, leads to the second broken symmetry. According to Eq. \ref{eq:exactome} the mean angular frequency reads:
\begin{align}
\langle \omega \rangle = &\frac{m(\hbar k \omega_T)^2}{2\eta^2} u (s_y-s_x) \sqrt{\frac{1-u^2}{4T_x T_y + u^2 (T_x -T_y)^2}}.
\end{align}
The scheme could not be implemented on standard alkaline atoms where broad transitions lead to Doppler temperatures usually higher than the potential depth. In contrast, narrow intercombination lines of Alkaline-earth atoms are favorable to such experiments. For instance, cooling of bosonic Strontium 88 on the intercombination line $^1S_0\rightarrow\,^3P_1$ of linewidth $\Gamma/2\pi=7.5 \text{kHz}$, leads to temperatures in the microKelvin range compatible with usual dipole trap depth \cite{KATORI1999-2,CHALONY2011}. For an illustrative and realistic example, we take a dipole trap frequency of $\omega_T = 250~\text{Hz}$, saturation parameters $\{\frac{I_x}{I_s},\frac{I_y}{I_s}\}=1,4$ and a detuning $\delta=-3$. We choose as a trap anisotropic $u=0.4$ . Our model gives $T_x=0.39~\mu$K, $T_y=0.72~\mu$K and a mean angular velocity of $15$ Hz. Additionally, the inverse quality factor (dimensionless viscosity) $\eta'$ is close to 2, leading to an overdamped dynamic. Figure \ref{fig:tof} displays a simple time-of-flight (TOF) experiment to visualize the effect. After stirring the atoms, we release them from the trap and following their ballistic expansion along $x$ and $y$. The clockwise (upper) and counter-clockwise (lower) cases clearly show a net mean rotation. Those figures were realized using a Cholesky decomposition of the covariance matrix $\Xi$, which gives access to $(x,y,v_x,v_y)$ for an arbitrary number of independent particles. Here this number of atoms is chosen reasonably high ($n=2\times 10^7$) for a clear reading of the figures. Initially small, the cloud will expand and maintain an asymmetric shape, as if the rotation was rigid. Nevertheless, we keep in mind two important facts: first the rotation is not strictly rigid due to the $\theta$-dependence in Eq. \ref{jtheta}. Second, our model is for independent particles and thus, the optical depth has to be low such that multiple scattering, which couple atoms, can be disregarded. Finally, we note that the rotation is done in the strong overdamped limit. Indeed the characteristic decay time of the velocity is given by $m/\eta$ which is in the millisecond range, namely much shorter than $|\langle \omega \rangle|^{-1}$.

\section{Conclusion}
We have shown that for a two-dimensional particle undergoing a stochastic motion with the two different temperatures along perpendicular axes, and
subjected to an external force deriving from a confining potential, the system evolves to a stationary state.  in which a permanent current is
present when the two principal axes of the confining potential do not coincide  with  the temperature axes. We finally proposed an experiment with  laser cooled atomic system for observing this phenomenon.

 \appendix

 \section{Method}\label{Method}
We first introduce the method allowing to obtain the complete solution of the two above models. Indeed, the models belong to the class of linear multivariate Fokker-Planck equations\cite{VanKampen1992,Lax1960}.
Let us denote $y$ a $r$-dimensional vector, the linear Fokker-Planck equation is given by
\begin{equation}
 \frac{\partial P(y,t)}{\partial t}=-\sum_{i,j}A_{ij}\frac{\partial y_j P(y,t)}{\partial y_i}+\frac{1}{2} B_{ij} \frac{\partial^2}{\partial y_i\partial y_j} P(y,t).
\end{equation}
The solutions of this linear Fokker-Planck equation is a Gaussian distribution
\begin{equation}
 P(y,t)=\sqrt{\frac{(2\pi)^r}{ Det(\Xi)}} \exp\left(-\frac{1}{2}(y-\langle y\rangle)^T\Xi^{-1}(y-\langle y\rangle)\right)
\end{equation}
where $\langle \rangle$ denote the average over the variable, $y^T$ the transpose of $y$ and $\Xi\equiv \Xi(t)$ is a time dependent $r\times r$ covariance matrix.
By taking the first and the second moment of the Fokker-equation, the covariance matrix satisfies the differential equation

\begin{equation}\label{eq:matrix}
 \frac{d\Xi}{dt}=A\Xi+\Xi A^T+B
\end{equation}
where $A$ and $B$ are $r\times r$ matrices with coefficients $A_{ij}$ and $B_{ij}$, respectively.

Note that for the two models defined above, the matrices $A$ and $B$ are symmetric. Moreover,
for the sake of simplicity, one first considers the stationary solution, where the stationary covariance matrix is denoted as $\Xi_s$
obeying to the algebraic equation
\begin{equation}\label{eq:matrix2}
 A\Xi+\Xi^T A=-B.
\end{equation}

\begin{acknowledgments}
Pascal Viot acknowledges Gleb Oshanin and Olivier Benichou for fruitful discussions and the School
of Physical and Mathematical   Sciences, NTU, Singapore where a part of this work was done. The authors acknowledge warm-heartedly Fr\'ederic
Chevy and Dominique Delande for their careful proofreading and advises.
\end{acknowledgments}

\end{document}